\documentclass[twocolumn,a4paper,amsmath,floatfix]{revtex4-1}
\usepackage[utf8]{inputenc}
\usepackage{graphicx}
\usepackage{amsmath}
\usepackage{amssymb}
\usepackage{color}

\usepackage{wrapfig}
\usepackage{float}
\usepackage{tikz}
\usepackage{listings}
\usepackage{algorithm}

\newcommand{\hex}{\mathrm{hex}}

\begin{document}

\title{Low-temperature crystal structures of the hard core square shoulder model}
\author{Alexander Gabri\"else}
\author{Hartmut L\"owen}
\author{Frank Smallenburg}
\affiliation{Institut f\"ur Theoretische Physik II: Weiche Materie, Heinrich-Heine-Universit\"at D\"usseldorf, Universit\"atsstr. 1,
40225 D\"usseldorf, Germany}

\begin{abstract}
In many cases, the stability of complex structures in colloidal systems is enhanced by a competition between different length scales. Inspired by recent experiments on nanoparticles coated with polymers, we use Monte Carlo simulations to explore the types of crystal structures that can form in a simple hard-core square shoulder model which explicitly incorporates two favored distances between the particles. To this end, we combine Monte Carlo-based crystal structure finding algorithms with free energies obtained using a mean-field cell theory approach, and draw phase diagrams for two different values of the square shoulder width as a function of the density and temperature. Moreover, we map out the zero-temperature phase diagram for a broad range of shoulder widths. Our results show the stability of a rich variety of crystal phases, such as body-centered orthogonal (BCO) lattices, not previously considered for the square shoulder model.
\end{abstract}

\maketitle

\section{Introduction}

In the past decades, the principles of designed colloidal self-assembly \cite{Frenkel,vogel2015advances} have been widely used to generate novel structures on the mesoscale by tailored interactions \cite{zhang2013general,Sprakel} and external stimuli \cite{Yethiraj_vanBlaaderen,HL_EPJST,grzelczak2010directed}. The colloidal building blocks in this framework can have either spherical \cite{Yodh,Palberg,HL_book} or asymmetric shape (for examples see Refs.\ \cite{AP,Vroege,APreview}), and can range in size from the micrometer scale down to a few nanometers, where their shape and size can be controlled with atomic precision \cite{ANIE:ANIE201609036}. Regardless of shape or size, these particles self-assemble due to a combination of thermal noise, mutual interactions, and external forces. In order to predict and control colloidal self-assembly, it is crucial to understand the equilibrium bulk phase diagram for a given colloidal interaction. 

This framework of self-assembly provides an effective route towards the creation of an amazing range of colloidal crystal structures by tuning the interactions between colloidal building blocks. Even in the seemingly simple case of monodisperse particles with spherically symmetric interactions, an impressive array of different structures have been both predicted theoretically and observed in experiments. In particular, it has been shown that soft repulsive interaction potentials can be tuned to favor e.g. open crystal lattices such as diamond \cite{watzlawek1999phase,likos2002exotic}, lattices with large unit cells such as A15 \cite{ziherl2001maximizing}, and even quasicrystals \cite{barkan2011stability,denton1998stability,dotera2014mosaic}. These predictions are supported by experimental observations on e.g. soft spherical polymers, micelles or dendrons \cite{sakya1997micellar, ungar2003giant, lee2014sphericity}, as well as polymer-coated nanoparticles \cite{hajiw2015evidence,goodfellow2013ordered,kuttner2016macromolecular}, which all demonstrate a rich crystal phase behavior \cite{boles2016self}. 

\begin{figure}
\centering
\begin{tabular}{cc}
\includegraphics[height=0.4 \linewidth]{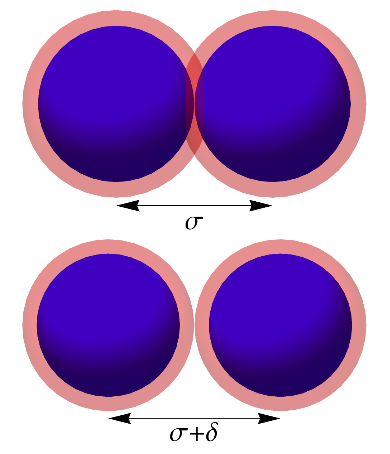} &\includegraphics[height=0.4\linewidth]{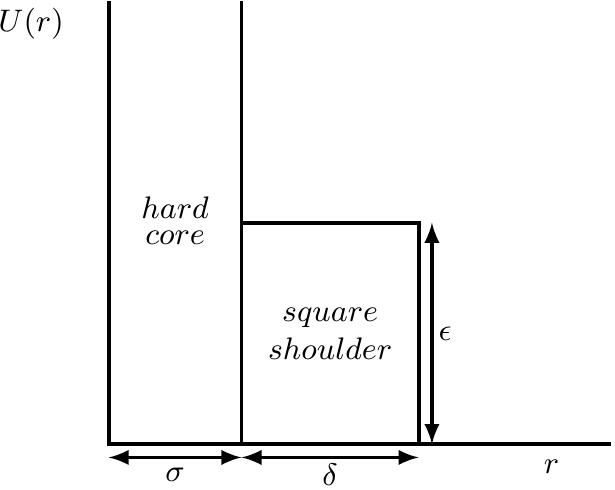}
\end{tabular}
\caption{Schematic representation of the hard core square shoulder (HCSS) model. The two images on the left show pairs of particles (blue) with their interaction range (pink), at distance $\sigma$ (top, energy $U = \epsilon$), and at distance $\sigma + \delta$ (bottom, energy $U = 0$). The interaction potential is plotted on the right.}
\label{fig:sq_potential}
\end{figure}

In many cases, the complexity of the structures that form in these systems can be understood from the presence of multiple favored length scales: certain particle separations are either favored or disfavored by the shape of the interaction potential. One of the most fundamental examples of such a model is the hard core square shoulder (HCSS) model, in which the spherical particles cannot overlap, and pay a fixed energy penalty for approaching each other too closely. In this case, the two length scales are set by the hard-core diameter $\sigma$ and the (larger) interaction range $\sigma + \delta$, as illustrated in Fig. \ref{fig:sq_potential}. Although the HCSS model is not designed quantitatively model a specific system, it can be seen as a phenomenological model for colloidal particles with a hard core and a soft corona, such as polymer-coated nanoparticles \cite{hajiw2015evidence,goodfellow2013ordered}. Moreover, it serves as a fundamental model for understanding the phase behavior of models incorporating a competition between two length scales, and as a result has received significant attention over the past decades. In two dimensions, such models have been shown to stabilize quasi-crystalline phases \cite{dotera2014mosaic,pattabhiraman2015stability,pattabhiraman2017phase}, as well as a large variety of crystal lattices \cite{malescio2003stripe,fornleitner2010pattern}. In three dimensions, fluid \cite{yuste2011structure} and glass states of the HCSS model have been shown to exhibit unusual behavior, such as polyamorphism and water-like anomalies \cite{buldyrev2009unusual,heyes1992square, bordin2017brownian,sperl2010disconnected}. Additionally, the crystal phase behavior of HCSS models has been studied extensively in the range of shoulder length $\delta/\sigma\in\{0.03,..,0.08\}$, where an isostructural transition between two face-centered cubic (FCC) structures occurs, ending in a critical point at high temperatures \cite{kincaid1976isostructural,bolhuis1997isostructural,denton1997isostructural}. In contrast, for long shoulder lengths $\delta/\sigma > 1.5$, particles can self-assemble into highly complex, low-symmetry lattices, forming clusters, columns, or lamellae \cite{pauschenwein2008clusters,pauschenwein2008zero}. In the intermediate regime, square-shoulder models have been predicted to form a number of crystal structures, including body-centered cubic (BCC) and A15 lattices \cite{rascon1997phase,ziherl2001maximizing}, both also observed in experiments of soft repulsive particles \cite{schmitt2016formation,li2004efficiency,imai2005static}. However, an exhaustive search of the range of crystal structures that might be stable in this regime at low temperatures is still lacking.

In this work, we systematically explore the ground-state phase behavior of the three-dimensional HCSS model over a broad range of interaction ranges, and identify a number of stable crystal structures which were not considered in earlier work. Additionally, we draw approximate phase diagrams using a mean-field cell theory for two choices of the interaction range, taking into account the free energies of both fluid and crystalline phases. Our results provide an excellent basis for future studies on e.g. the stability of quasicrystalline phases in three-dimensional HCSS models, as well as the development of more detailed models for e.g. soft-shelled nanoparticles.

The paper is organized as follows. In Sec. \ref{sec:methods} we describe the methods we use to identify candidate crystal structures and to determine the free energy of the fluid and crystal phases. The results are presented in Sec. \ref{sec:results}, where we map out the zero-temperature phase diagram for the HCSS model with shoulder lengths $0 < \delta / \sigma < 0.41$, and draw the phase diagrams in the density-temperature plane for $\delta/\sigma = 0.15$ and $0.2$. We present our conclusions and discussion in Sec. \ref{sec:conclusion}. 

\section{Model and simulation methods}
\label{sec:methods}

The square shoulder potential \cite{kincaid1976isostructural} between a pair of particles can be written as
\begin{equation}
U(r) =
  \begin{cases}
    \infty       & \quad r < \sigma\\
    \epsilon  & \quad \sigma < r < \sigma+\delta\\
    0		& \quad r > \sigma+\delta\\
  \end{cases},
\end{equation}
where $r$ is the distance between the particle centers, $\sigma$ the diameter of the particle and $\epsilon$ the potential of the shoulder with length $\delta$.

In order to study the phase behavior of this model system, we first predict candidate crystal structures using the ``floppy box'' Monte Carlo method \cite{filion2009efficient,de2012crystal}, and then determine the free energies of these structures using an approximate mean-field cell theory. Additionally, we use thermodynamic integration to determine the free energy of the fluid phase. From these free energies, we determine the phase diagrams for shoulder widths $\delta / \sigma = 0.15$ and $0.2$ using common tangent constructions. In the following subsections we describe the crystal prediction and free energy calculation methods in more detail.

\subsection{Crystal structure prediction} 
\label{sec:fbmc}

A crucial step in exploring the phase behavior of any colloidal model is the identification of crystal structures that should be taken into account as potentially stable phases. Although for very simple models one can often reasonably guess what crystals may be relevant, the prediction of stable structures from an interaction potential is generally far from straightforward. Hence, the search for potentially stable crystal structures is often done by a systematic numerical search, using e.g. genetic algorithms \cite{bianchi2012predicting,fornleitner2010pattern,gottwald2005predicting} or simulations of single unit cells \cite{filion2009efficient, de2012crystal, bianchi2012predicting}.

Here, we apply the floppy-box Monte Carlo (FBMC) method \cite{filion2009efficient, de2012crystal}. This method makes use of simulations of extremely small simulation cells with periodic boxes, at constant number of particles $N$, pressure $P$, and temperature $T$. The small number of particles in the box ($N \leq 14$) allows for rapid sampling of different crystal structures. The three vectors which span the simulation box, vary separately both in length and orientation. During the simulation, we slowly increase the temperature to quench the system into a low-energy state, and examine the resulting structures from a large number of simulations with different choices of $1 \le N \le 14$ and $1 \le P\sigma^3/\epsilon \le 20$. This approach is highly likely to find structures that are stable in the low-temperature limit, where at any given density, the structure or coexistence of structures with the lowest potential energy is the stable phase. Additionally, structures which are entropically favored are more likely to occur frequently \cite{filion2009efficient}. Hence, we look for structures that are either stable at zero temperature (i.e. have the lowest energy at a given density), or occur repeatedly in our simulations, and use these structures in our free-energy calculations. We identify the symmetry group of the observed crystals using the FINDSYM program \cite{stokes2005findsym}. 

In addition the original FBMC approach, we make use of a variation of the same technique which uses shifted boundary conditions, as described in the Supplementary Information. Together with the crystal structures found via these methods, we consider in our calculations the A15 \cite{ziherl2001maximizing} and C14 \cite{hajiw2015evidence} structures which have been observed in previous studies of hard-core soft-shell particles.

\subsection{Crystal free energies}
\label{sec:cell_theory}

To estimate the free energy of the crystal phases, we use an approximate mean-field cell theory \cite{prestipino2009zero,lennard1937critical}. In this approach, the free energy of each particle is calculated by assuming that all other particles are located exactly at their lattice sites. The partition function for the particle under consideration can then be calculated numerically by inserting this particle randomly at different positions into a sufficiently large volume $V_0$ surrounding its lattice position $\mathbf{r}_{0}$, and determining the energy of the particle at that position. Specifically, the partition function $Q_1$ for the particle is given by 
\begin{equation}
 Q_1 = \frac{V_0}{\Lambda^3} \left\langle \exp\left(-\beta\left[u(\mathbf{r})-\frac{1}{2}u(\mathbf{r}_{0})\right]\right)\right\rangle_{V_0},
\end{equation}
where $u(\mathbf{r})$ is the energy of the particle at position $\mathbf{r}$, $\Lambda$ is the thermal wavelength, and $\beta = 1/k_B T$ with $k_B$ Boltzmann's constant and $T$ the temperature. Subsequently, the free energy per particle of the crystal is obtained by averaging the single-particle free energy $F_1 = - k_B T \log Q_1$ for all particles in the crystal unit cell.

For many crystal structures, the details of the unit cell vary with density and temperature. For example, a body-centered tetragonal (BCT) structure is essentially a body-centered cubic (BCC) crystal compressed in one direction, such that the lattice spacing $c$ along one of the axes is different from the spacing $a$ along the other two. The ratio between the unit cell lengths $c/a$ depends on both the density and temperature, and hence a calculation of the free energy of this structure needs to take this into account. In other crystal structures, the lengths and directions of the vectors controlling the unit cell, and the positions of particles in the cell, may similarly vary. To address this, we need to minimize the crystal free energy with respect to all free parameters at each density and temperature. However, this is computationally expensive when the number of free parameters is large. To speed up this process, we make use of the observation that in systems of purely hard spheres, the single-particle partition function can be calculated fairly accurately by estimating that the particle can move freely in a polyhedral volume obtained by moving all of the faces of the particle's Voronoi cell inwards by a distance $\sigma / 2$ \cite{ziherl2001maximizing}. Although this approximation slightly underestimates the entropy of the particle, it allows for significantly faster calculations. We extend this approach here to the square shoulder model by approximately dividing the insertion volume up into polyhedral regions with different potential energies, and construct the total single-particle partition function as:
\begin{equation}
 Q_1 = \frac{1}{\Lambda^3} \sum_i V_i \exp\left(-\beta [ u_i - u(\mathbf{r}_0)] \right),
\end{equation}
with $u_i$ the potential energy in subvolume $i$. We construct the boundaries of all polyhedral subvolumes $V_i$ by dividing surfaces obtained by shifting faces of the central particle's Voronoi cell inwards by either $\sigma / 2$ or $(\sigma + \delta) / 2$, and calculate their volumes using the Voro++ library \cite{rycroft2009voro++}. For more details, see the Supplementary Information. When comparing the Voronoi approach to the insertion approach for identical choices of free parameters, we obtain good agreement, with differences on the order of $0.05 k_B T$. The effect of this approximation is of similar magnitude as the errors expected from the mean-field cell theory assumptions.

Note that these free energies are approximate, and hence the predicted phase boundaries are expected to deviate from the true phase diagram, especially at higher temperatures where entropy plays a stronger role. Nonetheless, cell theory has proven to be effective in predicting the phase behavior of systems with both hard and soft interactions \cite{vega1995solid, cottin1993cell}.

\subsubsection{Fluid free energy}
\label{sec:virial}

To determine the free energy of the fluid, we perform Monte Carlo simulations in the $NPT$ ensemble in a simulation box with fixed cubic shape, and a fixed number of particles $N = 343$, for a broad range of number densities $\rho = N/V$ and temperatures. We then determine the fluid free energy by using thermodynamic integration \cite{frenkel1997understanding} of the equation of state $P(\rho)$:
\begin{equation}
\frac{\beta F(\rho)}{N} = \frac{\beta F_{id}(\rho)}{N} + \int_{0}^{\rho} d \rho' \frac{\beta P(\rho') - \rho'}{\rho'^2},
\label{eq:energy_fluid}
\end{equation}
where $F_{id}(\rho)=Nk_{b}T(\log \rho \Lambda^3 - 1)$ is the ideal gas free energy. When fitting the equation of state, we improve accuracy at low densities by making use of the second virial coefficient $B_2$, which can be calculated analytically and is given by
\begin{equation}
B_{2}=-\frac{2\pi}{3}(\sigma+\delta)^{3}(e^{-\beta\epsilon}-1)+\frac{2\pi}{3}\sigma^{3}e^{-\beta\epsilon}.
\end{equation}

\section{Results}
\label{sec:results}

\begin{figure} 
\centering
    \includegraphics[width=0.95\linewidth]{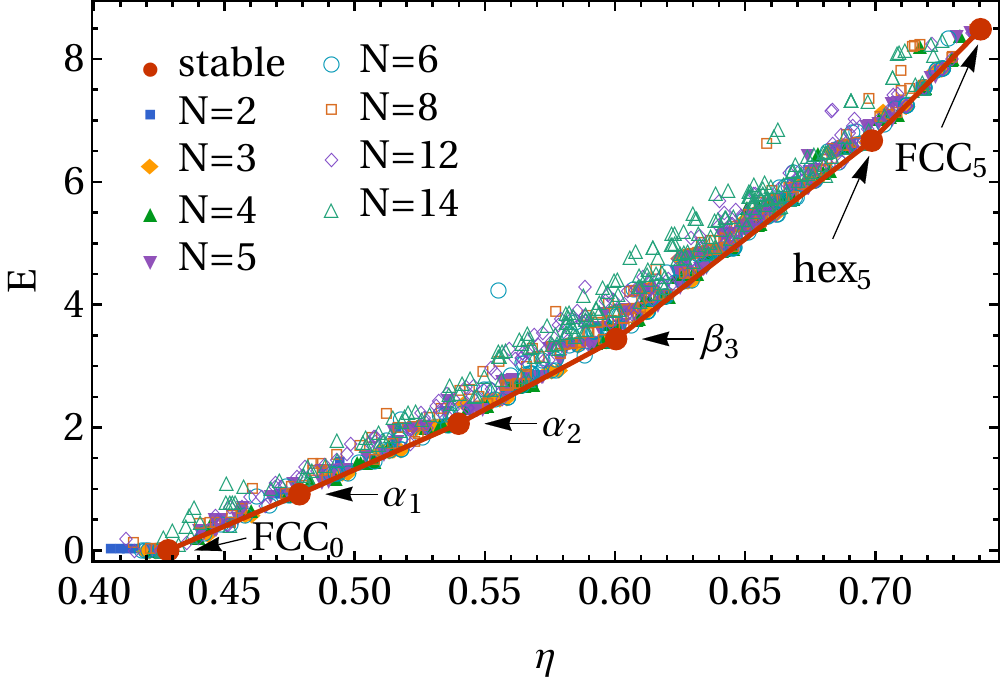}  
  \caption{Candidate crystal structures for interaction range $\delta / \sigma = 0.2$. Each point represents the density $\rho$ and potential energy per unit volume $E=\beta U\sigma^{3}/V$ of the final structure in an independent FBMC simulation. Simulations were run with varying choices for the number of particles $N$, temperature, and pressure. Note that for all stable structures, the same crystal was found from multiple independent simulations.} 
\label{fig:structuresd0p2}
\end{figure}

We begin our investigation by exploring the phase behavior of the square shoulder model in the zero-temperature limit. To this end, we use FBMC simulations to obtain candidate crystal structures for a range of choices of the interaction range $\delta$, and collect the dimensionless number density $\rho \sigma^3$ and potential energy per unit volume $E=\beta U\sigma^{3}/V$ of each structure for fixed $\delta$ in a diagram such as the one shown in Fig. \ref{fig:structuresd0p2}. In this representation, the potential energy of a coexisting state between two crystals is represented as a straight tie-line between the points corresponding to these crystals. Since at each density, the most stable state in the zero-temperature limit is the phase or coexistence of two phases with the lowest potential energy, the zero-temperature phase diagram can be obtained by simply connecting the lowest points at each density (effectively a common tangent construction), as shown in Fig. \ref{fig:structuresd0p2} for the interaction range $\delta/\sigma = 0.2$.

\begin{figure*}
\includegraphics[width=0.99\linewidth]{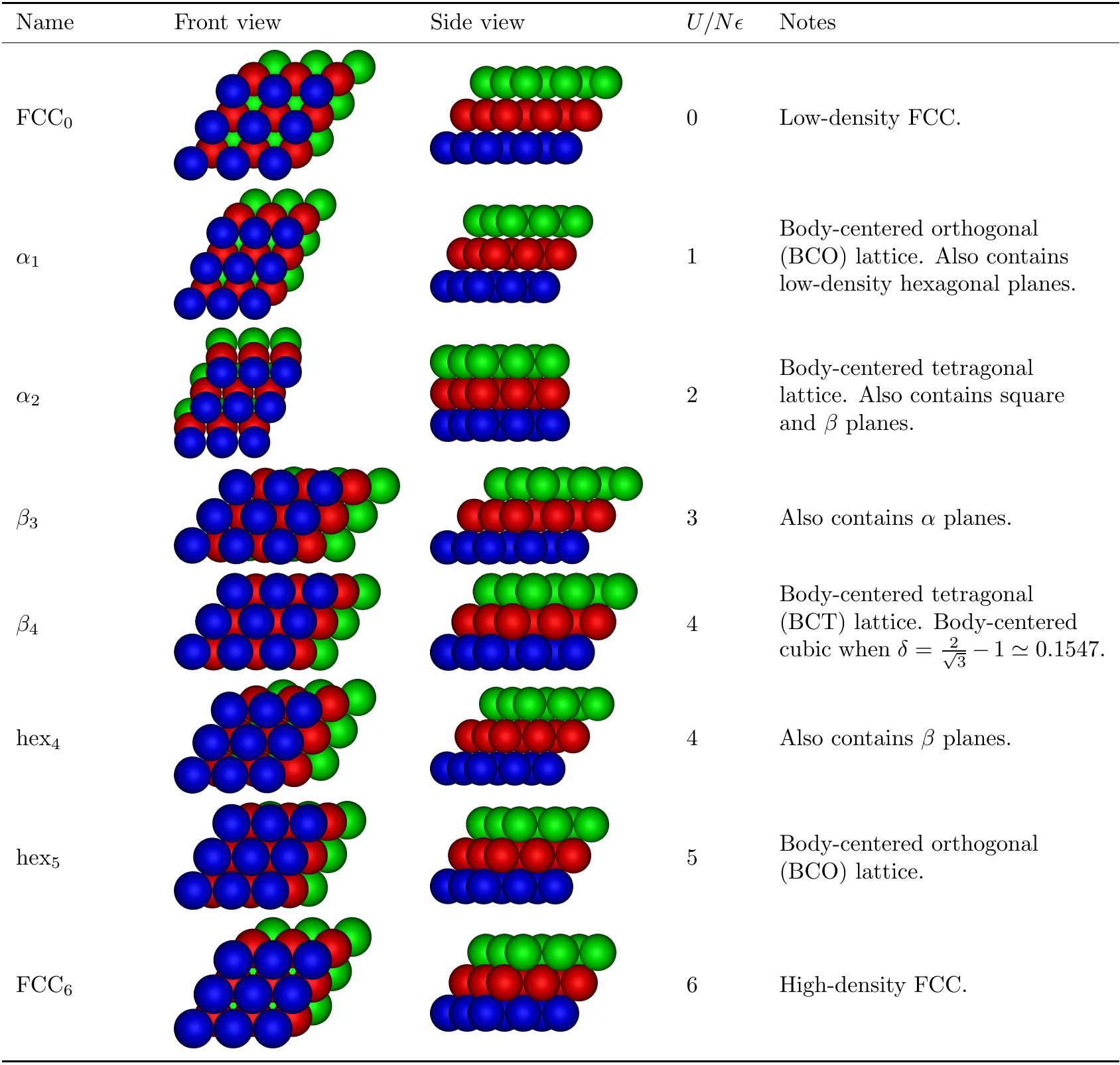}  
\caption{Graphical representations of stable ground-state crystal structures. For each crystal, we list a name, two snapshots with different planes highlighted in different colors, and any remarks on the details of the structure. In all snapshots, $\delta / \sigma = 0.2$, with the exception of $\beta_4$ (with $\delta / \sigma = 0.15$) and $\hex_4$ (with $\delta / \sigma = 0.25$).}\label{fig:structpics}
\end{figure*}

For this interaction range, we find six distinct stable crystal structures at zero temperature, differentiated by their potential energy per particle and density. With increasing density, we first encounter FCC$_0$, a low-density face-centered cubic phase (see Fig. \ref{fig:structpics}), with the subscript indicating the potential energy per particle $u = U/N\epsilon = 0$. This crystal consists of low-density hexagonal planes (lattice spacing $\sigma + \delta$), where particles interact neither with neighbors in the same plane, nor with neighbors in adjacent planes. Next, we obtain a crystal structure consisting of distorted hexagonal planes we call $\alpha$-planes, in which each particle interacts with two neighbors within the plane. These planes are stacked such that particles in neighboring planes do not interact (while maximizing the density), leading to a total energy per particle $u = 1$. The resulting crystal structure, which we call $\alpha_1$, is commensurate with a body-centered orthogonal (BCO) lattice. The next structure, $\alpha_2$, consists of the same planes, but has each particle interacting with an additional neighbor in each adjacent plane, leading body-centered tetragonal (BCT) lattice with a total of four bonds per particle ($u = 2$). Moving on to higher densities, we find a crystal structure consisting of $\beta$-planes: hexagonal planes distorted such that particles now interact with four in-plane neighbors. It should be noted that when these planes are stacked such that they form no out-of-plane bonds, we simply recover the $\alpha_2$ crystal in a different orientation. In the $\beta_3$ structures, the $\beta$-planes are instead stacked such that each particle interacts with one particle in each neighboring plane. Finally, in the high-density regime, we again find hexagonal planes, but this time with a lattice spacing of $\sigma$, such that each particle interacts with six neighbors within the same plane. These planes can again be stacked to allow for a different number of out-of-plane bonds, resulting in a body-centered orthogonal lattice $\hex_5$ with a total energy of $u = 5$, and finally $\hex_6$, corresponding to densely packed FCC, with total energy $u = 6$. 

It is important to note that these structures can all be modified by altering the stacking of consecutive layers. For example, while the simplest way of stacking hexagonal planes leads to an FCC structure, an alternate stacking choice leads to a hexagonally close-packed (HCP) crystal phase. In principle, an infinite number of distinct stackings is possible, all with the same potential energy and density. Similarly, for the other phases identified here, there are always multiple choices for placing consecutive layers, which correspond to different crystal structures. Since these are all equally stable at zero temperature, and our cell theory approach would not correctly capture free energy differences between them at finite temperatures, in this work we always assume the simplest possible stacking (corresponding to the smallest unit cell) in each case. Similarly, crystal structures containing identical planes can in principle be mixed by varying the way these planes are stacked. This essentially corresponds to a coexistence between the two phases. At zero temperature, there is no surface tension cost for these mixed stackings, and as such they are equally stable as a coexistence of the pure phases. At finite temperatures, however, we expect these mixed stackings to carry a finite interfacial free energy cost, and hence we only consider the pure phases for our phase diagrams.

We repeat our crystal search procedure for a broad range of interaction ranges $0 < \delta/\sigma = 0 < 0.41$. Note that $\delta/\sigma = \sqrt{2} - 1 \simeq 0.41$ marks the point where diagonal neighbors in a close-packed square arrangement start interacting. As a result, a number of the crystal structures predicted here are expected to change when $\delta$ is increased beyond this point. In the investigated regime, we found two additional stable structures, namely a $\beta_4$ structure corresponding to a BCT lattice, and a $\hex_4$ structure, both with an energy $u = 4$. 

In Fig. \ref{fig:structpics} we list all obtained structures, and provide snapshots from two different angles. Although we classify the structures by breaking them down into approximately hexagonal planes, several of them have additional symmetries which allow for a more precise identification of the structure. In particular, as listed in Fig. \ref{fig:structpics}, the $\alpha_1$ and $\hex_5$ crystals correspond to a body-centered orthogonal (BCO) lattice, and the $\alpha_2$ and $\beta_4$ crystals correspond to a BCT lattice. Interestingly, BCO lattices have been previously predicted for soft repulsive particles, such as star polymers \cite{watzlawek1999phase,likos2002interactions,hoffmann2004structure}.

\begin{figure}
\centering
    \includegraphics[width=0.95\linewidth]{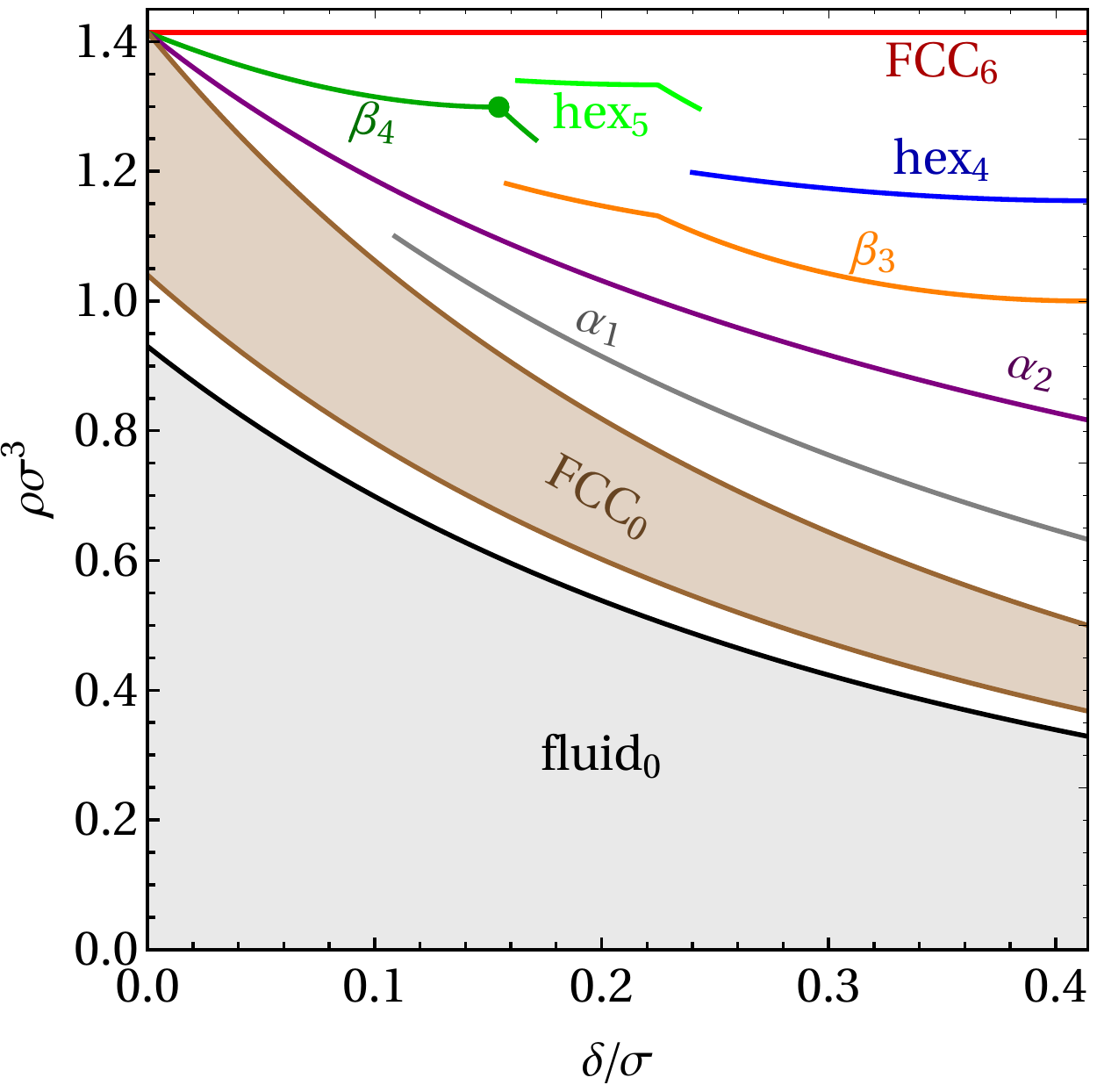}  
  \caption{Phase diagram of the HCSS model in the zero-temperature limit, as a function of interaction range $\delta$ and number density $\rho$. The white areas indicate coexistence regions. All tie lines between coexisting state points are vertical. The green point marks the value of $\delta/\sigma$ where the ground-state $\beta_4$ corresponds to a body-centered cubic (BCC) lattice. } 
\label{fig:zeroTphasediag}
\end{figure}

Combining our information on all stable structures, we systematically map out the phase diagram in the zero-temperature limit, for interaction ranges $0 < \delta / \sigma < 0.4$, and plot the result in Fig. \ref{fig:zeroTphasediag}. We also include here the stability regime of the fluid phase. At low densities, where packings with zero energy are possible, the system simply acts as a system of hard spheres of diameter $\sigma + \delta$, exhibiting a fluid-FCC coexistence between densities $0.94 < \rho(\sigma + \delta)^3 < 1.04$ \cite{noya2008determination}. Other crystal structures only show up beyond the maximum density for the low-density FCC phase ($\rho (\sigma + \delta)^3 = \sqrt{2}$). Note that at zero temperature, these higher-energy structures only appear at their maximum density, in order to allow as much of the system to remain in the lower-energy state as possible. This results in a phase diagram mostly filled by (white) coexistence regions, where the system is expected to macroscopically phase separate into the two adjacent phases. For example, at $\delta = 0.2$ and $\rho \sigma^3 = 0.57$ (located in the fluid-FCC coexistence region in Fig. \ref{fig:zeroTphasediag}), we expect the system to be divided into a fluid region at the freezing density of $\rho_f \sigma^3 = 0.54$ and a low-density FCC region at the melting density of $\rho_x \sigma^3 = 0.60$.

Even in the limit of zero temperature, the HCSS model shows a surprisingly rich phase diagram, with up to 7 different phases stable for a given $\delta$. Interestingly, many of these structures have not been considered in earlier studies exploring the phase behavior of square-shoulder models. All stable structures identified here can be represented by a unit cell containing only a single particle. Note that, apart from stacking variations, the FBMC simulations yielded no high frequencies of any structures beyond those stable at zero temperature, and the C14 and A15 lattices were not found to be stable in the zero-temperature limit.

\begin{figure}
\raggedright
{\bf a)} \vspace{-0.4cm}\\
\includegraphics[width=0.95\linewidth]{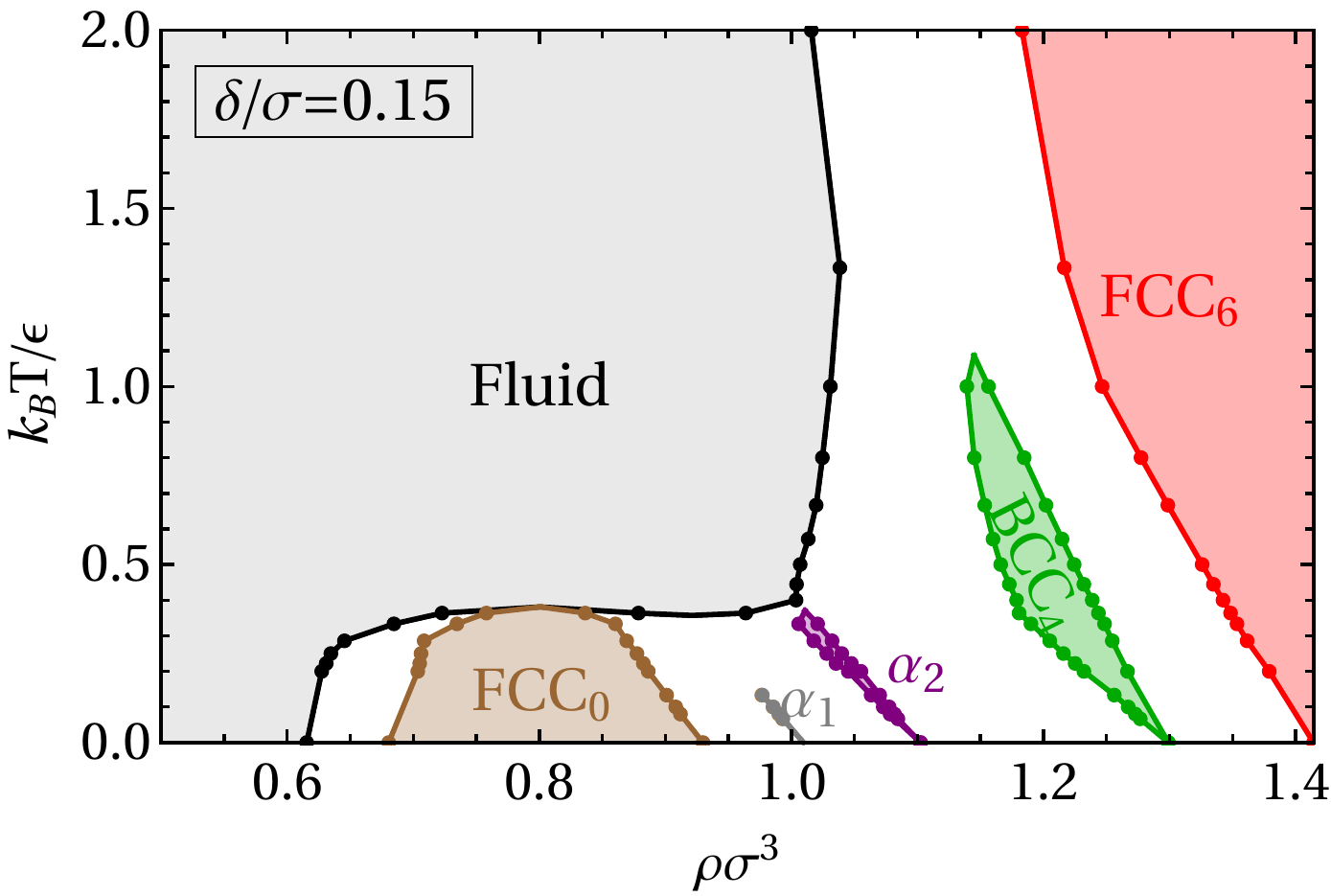} \\
{\bf b)} \vspace{-0.4cm}\\
\includegraphics[width=0.95\linewidth]{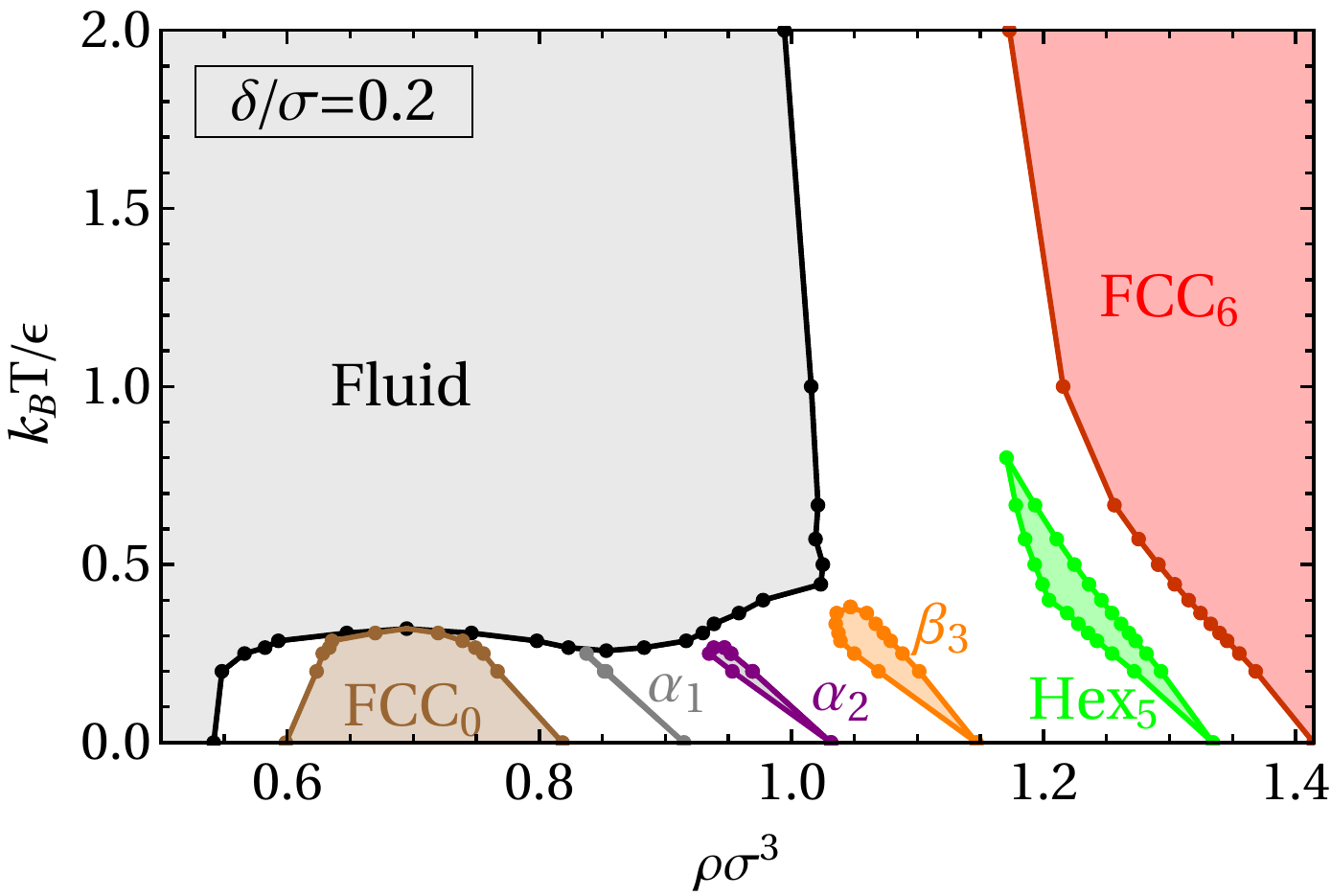}  
  \caption{Phase diagrams for interaction ranges $\delta/\sigma=0.15$ (a) and $0.2$ (b), as a function of density $\rho$ and temperature $T$. The white areas indicate coexistence regions. All tie lines between coexisting state points are horizontal. Coexistences (points) are obtained from free energies calculated using cell theory. Lines are guides to the eye only. Triple points are obtained by extrapolating coexistence lines.} 
\label{fig:phase_diagrams}
\end{figure}

While Fig. \ref{fig:zeroTphasediag} shows the structures which could in principle be found in the limit of strong interactions, the influence of entropy at finite temperatures can be expected to significantly change the phase behavior. To explore the effects of finite temperature, we focus on two values of the interaction range, $\delta / \sigma = 0.15$ and $0.2$, and map out the phase behavior as a function of temperature and density using cell theory. Here, we consider all crystal structures found in the zero-temperature limit, as well as the $C14$ and $A15$ lattices.

In Fig. \ref{fig:phase_diagrams}a we show the phase diagram in the density-temperature plane for interaction range $\delta=0.15$, as obtained from the free energies calculated using cell theory. Along the bottom axis, we recover the zero-temperature limit, showing a succession of phases including the fluid, low-density FCC$_0$, $\alpha_1$, $\alpha_2$, $\beta_4$, and high-density FCC$_6$ phases. Note that the $\beta_4$ rapidly and continuously transforms into a BCC crystal at finite temperatures, and is therefore labeled as BCC in the phase diagram. 
Although the $\alpha_1$ and $\alpha_2$ structures show only small regions of stability, both FCC phases and the BCC phase are stable in significant parts of the phase diagram. Interestingly, the fluid phase shows reentrant behavior close to the temperature where the low-density FCC phase vanishes. This results in a small regime where the FCC phase can melt upon both compression and expansion. 
With increasing temperature, the density range in which this occurs narrows and eventually vanishes at a single point, beyond which the low-density FCC phase is no longer stable. At sufficiently high temperatures, we recover the simple fluid-FCC coexistence expected in the high-temperature limit, which corresponds to the pure hard-sphere model.

We show the phase diagram for interaction range $\delta / \sigma = 0.2$ in Fig. \ref{fig:phase_diagrams}b. We observe similar phase behavior, where again only the phases stable at zero temperature occur in the phase diagram. In contrast to the findings in Ref. \cite{ziherl2001maximizing}, we find no stable BCC or A15 phase, as both phases are metastable with respect to melting. This discrepancy can likely be attributed to the fact that in Ref. \cite{ziherl2001maximizing}, the stability of the fluid phase at non-zero temperatures was estimated by assuming that the freezing density remained the same as in the zero-temperature limit. Our results thus highlight the importance of considering the enhanced stability of the fluid at non-zero temperatures. However, we cannot exclude the possibility of a stable A15 phase at larger $\delta$, where Ref. \cite{ziherl2001maximizing} predicts a larger region of stability for this structure.

\section{Conclusion}
\label{sec:conclusion}

We systematically explored the low-temperature phase behavior of the hard-core square-shoulder model, identifying a rich variety of distinct crystal structures which are stable in the zero-temperature limit. Additionally, we constructed phase diagrams in the density-temperature plane for interaction ranges $\delta/\sigma = 0.15$ and $0.2$ using cell theory, in order to estimate the temperature regimes in which these structures may be observed. Although all crystal structures identified here are relatively simple, consisting of unit cells containing only a single particle, their importance cannot be intuitively predicted a priori, illustrating the importance of a systematic search for unexpected stable structures.

Our results highlight the unexpectedly complex phase behavior that can result from the fundamental HCSS model. Moreover, it demonstrates that a small change in interaction range from $\delta/\sigma = 0.15$ to 0.2 can both stabilize and destabilize several crystal structures. This sensitivity also implies that small variations in the interaction potential shape, such as smoothening the sharp potential jump associated with the shoulder, can be expected to have far-reaching effects. Hence, it is not surprising that more realistic models for hard-core soft-shell particles have predicted a variety of crystal structures not seen here, including e.g. C14 and A15 phases \cite{pansu2017metallurgy}. Such phases are typically explained in terms of a minimization of the surface energy between building blocks \cite{ziherl2001maximizing}, resulting from e.g. the elasticity of the particle surface layers. In contrast, in the HCSS model, phases are stabilized purely due to the geometric interplay between the hard-core and soft-shoulder length scales, which leads to the stabilization of a distinct set of crystal structures. While this model potential is not expected to be quantitatively accurate for any specific colloidal system, it provides a framework for understanding the crystallization of colloids which interact differently at two length scales. However, quantitative predictions for the phase behavior of real colloidal systems will require carefully tuned model potentials, tailored to the details of the experimental system under consideration.

Although all predicted structures have small unit cells, it should be noted that our approach cannot exclude the possibility of stable crystals with larger unit cells than those considered in our FBMC simulations, or the possibility of stable quasicrystals, which are not commensurate with a periodic unit cell. In particular, quasicrystalline layers have recently been shown to form in a sedimenting HCSS system with shoulder length $\delta / \sigma = 0.4$ \cite{pattabhiraman2017periodic}. By providing a clear overview of competing periodic structures, the phase behavior predicted here provides important guidelines for exploring the stability of quasicrystalline or other non-periodic structures in this fundamental model.

As an outlook we emphasize that the rich equilibrium phase diagram obtained here for the square-shoulder model provides the starting point for future studies of guided colloidal self-assembly. The variety of stable crystal structures presented in this work may constitute important building blocks for photonic \cite{photonic1,photonic2} and phononic \cite{phononic1,phononic2} crystals with unusual material properties. Several techniques are conceivable to produce these exotic structures, including colloidal templating \cite{vanBlaaderen,Velikov} and field-directed crystallization \cite{Furst}. In the first case, a stable but kinetically blocked crystalline structure is forced to occur by an external template that incorporates the symmetry of the final desired equilibirum crystal structure. In the second case, a time-dependent external field produces dynamical channels to force the system to relax into the desired crystal. For both approaches, a detailed understanding of the equilibrium phase behavior, and the presence competing crystal structures, is crucial. Hence, the phase diagrams and crystal structures predicted here may pave the way towards the design of new colloidal crystal structures for various applications.

\medskip

We gratefully acknowledge support from the  Deutsche  Forschungsgemeinschaft  (grant  LO 418/19-1).

\bibliographystyle{prsty}
\bibliography{ref}

\end{document}